\documentclass[aps,prc,singlecolumn,superscriptaddress]{revtex4-2}

\usepackage{graphicx}

\usepackage{bm}

\usepackage{color}     
\usepackage{amsmath}
\usepackage{booktabs}
\usepackage{multirow}
\usepackage{array}
\usepackage{xcolor}
\usepackage{xpatch}

\usepackage{tikz,xcolor}
\usepackage{ulem}

\begin{document}

\title{Microscopic nuclear structure study of $^{229}$Th by Projected Shell Model}

\author{Zi-Rui \surname{Chen}}%
\affiliation{School of Physics, Nankai University, Tianjin 300071, China}

\author{Long-Jun \surname{Wang}}
\email{longjun@swu.edu.cn}
\affiliation{School of Physical Science and Technology, Southwest University, Chongqing 400715, China}

\author{Yuanbin \surname{Wu}}%
\email{yuanbin@nankai.edu.cn}
\affiliation{School of Physics, Nankai University, Tianjin 300071, China}

\date{\today}


\begin{abstract}

$^{229}$Th, a crucial candidate for nuclear clocks and many other applications, is a typical heavy nucleus with an extremely low-energy isomeric state $^{229m}$Th. A detailed study of the nuclear structure of $^{229}$Th is performed here by the microscopic model of state-of-the-art projected shell model. Our calculation describes well low-energy levels of $^{229}$Th, and provides a reduced transition probability $B(M1)$ of $0.0240$ W.u. for the isomeric transition which agrees well with the radiative lifetime of $^{229m}$Th measured recently. Our result supports a small multipole mixing for the cross-band transition of the second-excited state of $^{229}$Th, suggesting that further investigations on the inconsistencies in the decay of the second-excited state should be necessary. The physics behind these properties is revealed by the analysis of the nuclear wave functions. Our findings provide a deep insight into $^{229}$Th from the microscopic nuclear structure point of view, and offer the chance for further studies for nuclear clocks and relevant topics by microscopic nuclear structure theory.

\end{abstract}

\maketitle


\section{Introduction}

The $^{229}$Th isotope is known as a crucial candidate for nuclear clocks which may exceed the present accuracy of atomic optical clocks and be less sensitive to environmental conditions as compared with the atomic ones \cite{peik_2003_EPL,rellergert_2010_PRL,von_2020_EPJA,beeks_2021_natureRP,campbell_2012_PRL,kazakov_2012_NJP,peik_2021_QST}. This is mainly because of the unique isomeric state $^{229m}$Th, which is the lowest known metastable nuclear excited state \cite{walker_1999_nature,walker_2020_PS} with an extremely low excitation energy about $8$ eV. In addition to a possible nuclear clock, $^{229}$Th has also been shown to be a fascinating candidate in studies of fundamental physics \cite{peik_2021_QST,Flambaum_2006_PRL,berengut_2009_PRL,flambaum_2016_PRL,fadeev_2020_PRA,fadeev_2022_PRC,thirolf_2019_AP,shabaev_2022_prl} and applications \cite{beeks_2021_natureRP,tkalya_2011_PRL}, such as the detection of temporal variations of fundamental constants and dark mater, studies of nuclear hyperfine mixing, and nuclear gamma-ray lasers.

Numerous efforts have been made to populate the isomeric state $^{229m}$Th and measure relevant properties. Recently, direct laser excitation to the isomeric state in Th-doped crystals \cite{tiedau_2024_PRL,elwell2024laser,zhangck_2024_nature}, which is a key for a nuclear clock, has been demonstrated, after a number of efforts to populate $^{229m}$Th by various methods \cite{beck_2007_PRL, von_2016_nature, thielking_2018_nature, Seiferle_2017_PRL, benedict_2019_nature, Yamaguchi_2019_PRL, Sikorsky_2020_PRL, yamaguchi_2024_nature, kraemer_2023_nature, masuda_2019_nature,Hiraki_2024_NC}. The excitation energy of $^{229m}$Th is the key for the direct laser excitation \cite{von_2017_PRL}, which has been measured in several ways \cite{benedict_2019_nature, beck_2007_PRL, Yamaguchi_2019_PRL, Sikorsky_2020_PRL, kraemer_2023_nature, tiedau_2024_PRL,elwell2024laser,zhangck_2024_nature}. Electromagnetic moments in the ground and isomeric state have also been studied via laser spectroscopy \cite{thielking_2018_nature,campbell_2011_PRL,muller_2018_PRA,safronova_2013_PRA,zhangck_2024_nature}. The lifetimes and electromagnetic transition properties for higher levels have been measured \cite{kroger_1976_NPA, canty_1977_JPG, bemis_1988_ps, gulda_2002_npa, barci_2003_PRC, ruchowska_2006_PRC} in early years. The transition properties for the second-excited state of $^{229}$Th have also been measured and discussed in Refs.~\cite{masuda_2019_nature, Sikorsky_2020_PRL,nikolay_2022_PRC}. Recently, the radiative lifetime of $^{229m}$Th has been measured successfully, via the photon emission from $^{229m}$Th \cite{zhao_2012_PRL, kraemer_2023_nature, tiedau_2024_PRL, elwell2024laser,zhangck_2024_nature,Hiraki_2024_NC} or the laser spectroscopy of $^{229m}$Th$^{3+}$ \cite{yamaguchi_2024_nature}.

So far, the understanding of this unusually low-energy isomer and its decay properties from the nuclear structure point of view is limited. Nuclear structure studies of $^{229}$Th could provide not only supports for the development of a possible nuclear clock but also a deep understanding of nuclear structure theory via this typical and unique representative of heavy nuclei as a probe \cite{nikolay_2017_PRL, nikolay_2019_PRL, minkov_2021_PRC, Minkov_PRC_2024}. So far, the properties of $^{229m}$Th have been studied theoretically via the indirect method such as Alaga rules \cite{dykhne_1998_jetp, tkalya_2015_PRC}, and collective-based nuclear structure models, including the rotational model \cite{barci_2003_PRC}, the quasiparticle-plus-phonon model \cite{ruchowska_2006_PRC}, and sophisticated model based on the collective quadrupole-octupole vibration-rotation motion of the nucleus \cite{nikolay_2017_PRL, nikolay_2019_PRL, minkov_2021_PRC}. It should be highly demanded to employ microscopic nuclear models that account for the subtle interplay between single-particle and collective degrees of freedom in such a nucleus.

For the description of the nuclear system $^{229}$Th, we adopt a microscopic model of the projected shell model (PSM) \cite{PSM_review, PSM-Sun}, which relies on the mechanism of breaking of the rotational, reflection and U(1) (particle number) symmetries in the intrinsic nuclear states. In such a way many relevant correlations are already accounted for in the multi-quasiparticle (qp) basis employed. However, such intrinsic states, break several fundamental symmetries of the considered effective Hamiltonian and, therefore, such symmetries must be recovered a posteriori via the use of symmetry-projected configuration mixing techniques. In this way the PSM builds an effectively truncated, highly correlated and symmetry-projected multi-quasiparticle basis subsequently used to diagonalize the nuclear Hamiltonian, and the resulting (laboratory frame) wave functions can be used to access spectroscopic properties of the considered system. The PSM has been applied in studies of nuclear structures of deformed heavy nuclei \cite{cui_2014_PRC, cui_2012_cpc, chen_2008_prc, Devi_2005_PRC, bhat_2012_EPJA, Ram_2011_PDAESNP, Ram_2018_AIPCP, Sehgal_2006_JPG, LJWang_2016_PRC, vargas_2024_EPJA, velazquez_1999_NPA, sandoval_2011_PRC, wen_1996_PRC, Wu_2017_PRC}, novel vibration modes \cite{sun_1998_PRL, Lv_2022_PRL}, and weak processes \cite{Z_C_Gao_2006_GT, LJWang_2021_PRL, wangyk_2021_PRC, YXiao_2024_PRC, ZRC_2024_PLB}. Here, we develop state-of-the-art PSM to provide a natural description of heavy odd-mass nuclei, with large model and configuration spaces, octupole degree of freedom and corresponding parity projection, as well as the involvement of most of the important interactions for the first time. Our calculation provides a reduced transition probability $B(M1)$ of $0.0240$ W.u. (Weisskopf units) for the isomeric transition, which agrees well with the radiative lifetime of $^{229m}$Th measured recently \cite{kraemer_2023_nature, tiedau_2024_PRL, yamaguchi_2024_nature, elwell2024laser, Hiraki_2024_NC, zhangck_2024_nature}. Our calculation supports a small multipole mixing for the cross-band transition of the second-excited state of $^{229}$Th \cite{nikolay_2022_PRC}. By analyzing the nuclear wave functions, we also reveal the physics behind these results. Our findings offer an insight into this unique $^{229}$Th from the microscopic nuclear structure point of view, supporting further studies for nuclear clocks and relevant topics.

\section{Theoretical approach}

Our PSM calculations start from deformed Nilsson mean-field with pairing correlations by the Bardeen-Cooper-Schrieffer (BCS) method where the average neutron and proton numbers are constrained to those of $^{229}$Th, from which different orders of many-body configurations are constructed. Specifically, for $^{229}$Th, the configuration space is adopted as,
\begin{equation}
  \label{qpbasis1}
  \{ \hat{a}^{\dag}_{n_{i}}|\Phi (\varepsilon )\rangle,\ \hat{a}^{\dag}_{n_{i}}\hat{a}^{\dag}_{n_{j}}\hat{a}^{\dag}_{n_{k}}|\Phi (\varepsilon )\rangle,\ \hat{a}^{\dag}_{n_{i}}\hat{a}^{\dag}_{p_{j}}\hat{a}^{\dag}_{p_{k}}|\Phi (\varepsilon )\rangle \}, 
\end{equation}
where $|\Phi(\varepsilon )\rangle$ is the qp vacuum associated with the intrinsic deformation $\varepsilon$, and $\hat{a}^{\dag}_{n}(\hat{a}_{p}^{\dag})$ labels neutron (proton) qp creation operator. In this work, three major shells with $N = 5, 6, 7$ ($N = 4, 5, 6$) for neutron (proton) are adopted for the model space. Up to 3-qp configurations are considered, as the contribution from 5-qp configurations is expected to be small according to our calculations (see Appendix A). The intrinsic axial deformation is described by the PSM with parameters of quadrupole $\varepsilon_{2} = 0.19$, octupole $\varepsilon_{3} = 0.066$, and hexadecapole $\varepsilon_{4} = -0.089$ which are based on Refs.~\cite{Moller_1995, Moller_2016, nikolay_2017_PRL,Nomura_PRC_2014, Agbemava_PRC_2016}. The Nilsson parameters \cite{greiner_book} with the constant $\kappa$ of the spin-orbit term and $\mu$ of the orbit square term are presented in Table \ref{Tab:nilsson}, which are taken from Ref. \cite{para1985} with only the neutron $N=6$ parameters fitted slightly from $\kappa=0.062$ and $\mu=0.34$. The $N=6$ valence neutron Nilsson parameters used in this work are slightly adjusted from the original values optimized for actinide nuclei with specific quadrupole and hexadecapole deformations \cite{para1985}. This recalibration is necessitated by the distinct nuclear deformation characteristics of $^{229}$Th, realizing accurate reproduction of band-head energies (especially for the isomeric state).

\begin{table}[!htbp]
 
  \centering
  \renewcommand\arraystretch{1.2}

  \setlength{\tabcolsep}{0.7mm}{
  \begin{tabular}{cccccccccccccccc}
  \toprule[1pt] 
 
   & & & \multicolumn{5}{c}{Neutron}  & & & & \multicolumn{5}{c}{Proton} \\
   $N$ & & & $5$ & & $6$ & & $7$ & & & & $4$ & & $5$ & & $6$ \\
  \cline{4-8} 
  \cline{12-16}

  $\kappa$ & & & $0.062$ & & $0.052$ & & $0.062$ & & & & $0.065$ & & $0.060$ & & $0.054$\\
  $\mu$      & & & $0.430$ & & $0.230$ & & $0.260$ & & & & $0.570$ & &$0.650$ & & $0.690$\\

  \bottomrule[1pt] 
  \end{tabular}
  }
  \caption{Nilsson parameters adopted in the present work.}
  \label{Tab:nilsson}
    
  \end{table}

The many-body configurations in Eq. (\ref{qpbasis1}) including octupole deformation break the rotational and reflection symmetries, which can be restored exactly by the angular-momentum projection (AMP) \cite{PSM_review} and parity projection (PP) \cite{egido_1991_npa} operators,
\begin{eqnarray}
  \label{projected}
  \hat{P}_{MK}^{J} &=& \frac{2J + 1}{8\pi^{2}}\int d\Omega D_{MK}^{J} (\Omega)\hat{R}(\Omega), \\
  \hat P^\pi  &=& \frac{1}{2}(1+\pi\hat{\Pi}),
\end{eqnarray}
where $\hat{R}$ and $D_{MK}^{J}$ (with Euler angle $\Omega$ \cite{Angular_book}) are the rotation operator and Wigner $D$-function respectively, with $K$ ($M$) being the projection of the angular momentum $J$ in the intrinsic (laboratory) frame. $\pi = \pm 1$ labels the parity and $\hat{\Pi}$ is the parity operator. Particle number projection is neglected. The involvement of particle number projection remains computationally demanding for large configuration models. However, the neglect of particle number projection should not alter the fundamental physical picture established in the previous work of PSM~\cite{PSM_review}. Moreover, the effects of particle number fluctuation are expected to be small~\cite{Yao_2009_PRC, Yao_2010_PRC, Olofsson_PRA_2007}. Thus, in the present work, particle number conservation is treated only on an average basis through first-order Lagrange multipliers as in previous works~\cite{ cui_2014_PRC, cui_2012_cpc, chen_2008_prc, Devi_2005_PRC, bhat_2012_EPJA, Ram_2011_PDAESNP, Ram_2018_AIPCP, Sehgal_2006_JPG, LJWang_2016_PRC, vargas_2024_EPJA, velazquez_1999_NPA, sandoval_2011_PRC, wen_1996_PRC, Wu_2017_PRC, WangYK_2019_PRC, WangYK_2024_PLB, WangYK_2020_PLB, Yao_2009_PRC, Yao_2010_PRC}. The many-body wave function in the laboratory frame can then be described as, 
\begin{align} \label{wave_function}
     \Big|\Psi^{\xi}_{J^{\pi}M} \Big\rangle = \sum_{K \rho} f_{K\rho}^{J^\pi \xi} \  \hat P^{\pi}\hat{P}_{MK}^{J}|\Phi_{\rho}(\varepsilon )\rangle,
\end{align}
where $|\Phi_{\rho}(\varepsilon )\rangle$ represents the qp configurations in Eq. (\ref{qpbasis1}). $f_{K\rho}^{J^\pi \xi}$ denotes the expansion coefficients for the $\xi$-th eigen value with spin-parity $J^\pi$, which can be obtained by solving the Hill-Wheeler equation with the PSM Hamiltonian \cite{ButlerRMP1996}
\begin{align} \label{eq.Hamil}
  \hat H = \hat H_0 &  - \frac{1}{2} \sum_{L=2}^{4}\chi_{L}\sum_{\omega = -L}^{L}\hat{Q}^{\dag}_{L\omega} \hat{Q}_{L\omega}  \nonumber \\
                     &  - G_{M}\hat{P}^{\dag}\hat{P}-G_{Q}\sum_{\omega = -2}^{2}\hat{P}_{2\omega}^{\dag}\hat{P}_{2\omega},
\end{align}
where $\hat{H}_{0}$ is the single-particle part. The rest includes the quadrupole-quadrupole, octupole-octupole, hexadecapole-hexadecapole, monopole-pairing and quadrupole-pairing interactions, where $\hat{P}^{\dag}$ is the nucleon-pair creation operator, $\hat{Q}_{L\omega}$ and $\hat{P}_{2\omega}$ correspond to the multipole and quadrupole operators in particle-hole and particle-particle channels respectively~\cite{PSM_review}. $\chi_{L}$ is determined by self-consistent calculations of the deformed mean field \cite{ PSM_Sun2, PSM_review, Gao_PhysRevC_2000}.The monopole-pairing strength has the form $G_{M} = [G_{1} \mp G_{2}(N_n-Z)/A]/A$, where $+$($-$) in $\mp$ is for protons (neutrons), and $N_n$ , $Z$ and $A$ are the neutron, proton and mass number, respectively. The pairing strength parameters adopted in this work $G_{1} = 16.32$, $G_{2} = 11.81$ are slightly modified from the reference values optimized for even-even thorium isotopes $^{230, 232}$Th \cite{cui_2012_cpc}. $G_{Q}/G_{M} = 0.13$ is taken from the value for nuclei in this mass region \cite{cui_2012_cpc, chen_2008_prc}. With the wave function (\ref{wave_function}), the reduced electromagnetic transition strength can be calculated by \cite{Suhonen_book}, 

\begin{align} \label{eq.ROBTD}
  \left\langle \Psi_{J'^{\pi'}}^{\xi'} \big\|  \hat{\mathcal T}^\lambda_\sigma \big\|   \Psi_{J^\pi}^\xi  \right\rangle 
     =& \frac{1}{\sqrt{2\lambda+1}}
       \sum_{\alpha\beta} \left\langle \alpha \big\| \hat{\mathcal T}^{\lambda} \big\| \beta \right\rangle \nonumber\\
     &  \times  
    \left\langle \Psi_{J'^{\pi'}}^{\xi'} \big\|  [\hat{c}^{\dag}_{\alpha} \otimes \tilde{\hat{c}}_{\beta} ]^{\lambda}  \big\| \Psi_{J^\pi}^\xi  \right\rangle,
\end{align}
where $\hat{\mathcal T}^\lambda_\sigma$ is the electromagnetic transition operator \cite{Suhonen_book, Ring_book_2004}, $|\alpha\rangle \equiv | n_{\alpha}, l_{\alpha}, j_{\alpha}\rangle$ is the spherical basis, $\hat{c}^{\dag}$ and $\tilde{\hat{c}}$ are the single-particle creation and annihilation operators in the form of irreducible spherical tensor, which are associated with the qp basis by the Hatree-Fock-Bogoliubov transformation. $\left\langle \Psi_{J'^{\pi'}}^{\xi'}\big\|  [\hat{c}^{\dag}_{\alpha} \otimes \tilde{\hat{c}}_{\beta} ]^{\lambda}  \big\| \Psi_{J^\pi}^\xi \right\rangle$ represents the reduced one-body transition density (ROBTD) which can be calculated by the Pfaffian algorithm as in Refs. \cite{ZRChen_PRC, FGao_PRC2023}.

\section{Numerical results and discussions}

\begin{figure*}[!htbp]
\begin{center}
  \includegraphics[width=0.95\textwidth]{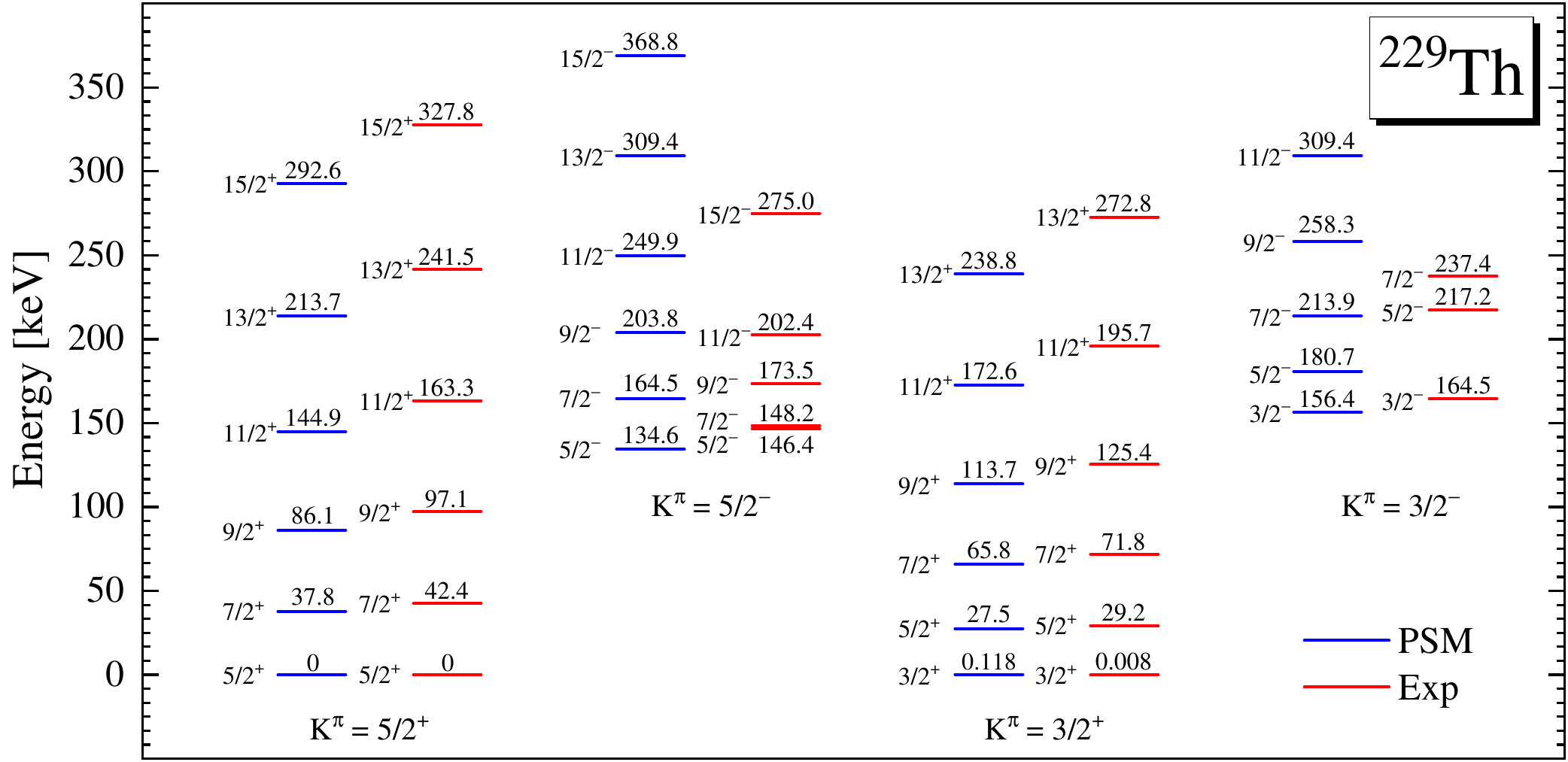}
  \caption{\label{fig:fig_1} Low-energy states of $^{229}$Th nucleus from the PSM calculation in the present work and the experimental data \cite{benedict_2019_nature,NNDC}.} 
\end{center}
\end{figure*}

Figure \ref{fig:fig_1} displays the calculated low-energy level structure of $^{229}$Th as compared with the experimental data \cite{benedict_2019_nature, NNDC}. It is shown that the ground state and isomeric state are reproduced, the yrast band based on the ground state and the isomer band as well as the two low-lying negative-parity bands are well described. From the mean-field calculations we found that the neutron Fermi surface locates at the 1-qp state 3/2 [631], while the 5/2 [633] state is higher for which the qp energy is 75 keV larger than the 3/2 [631] state. After projections and configuration mixing due to the Hamiltonian (especially the quadrupole-pairing term), the 5/2 [633] state becomes the main configuration of the ground state while the 3/2 [631] state becomes the main configuration of the very low-lying isomer. We also found in our calculations that the isomer energy is sensitive to the neutron $N=6$ single-particle levels and the pairing strengths. Furthermore, the $5/2^-$ ($3/2^-$) level is found to have the same main configuration as the ground state (isomer), but opposite parity caused by the exact parity projection.

Both our calculation and the model in Ref.~\cite{nikolay_2017_PRL} demonstrate the rotational characteristics in the positive-parity bands. While the angular momentum projection severs as the primary framework for nuclear rotation, the large deformed nuclear configuration incorporates quadrupole, octupole, and hexadecapole deformations with corresponding multipole interactions in the Hamiltonian, also enables the description of vibrational modes to some degree. This approach shares conceptual similarities with $\gamma$-vibrational studies \cite{Sheikh_PRC2011}, and consequently provides a band structure that closely matches the two negative-parity bands described by the collective quadrupole-octupole vibration-rotation model in Ref.~\cite{nikolay_2017_PRL}. Both our calculation and the calculation in Ref.~\cite{nikolay_2017_PRL} describe in a similar way the vibration mode above $7/2^{-}$ level in $K^{\pi}=5/2^{-}$ band and rotational structure characteristics in $K^{\pi}=3/2^{-}$ band. It is shown in the experimental data that the $5/2^{-}$ and $7/2^{-}$ levels are close (very close) in energy in the $K^{\pi}=3/2^{-}$ ($K^{\pi}=5/2^{-}$) band; however both our present work and the model in Ref.~\cite{nikolay_2017_PRL} cannot precisely describe this behaviour. This implies that further theoretical and experimental investigations on these two negative-parity bands should be necessary for the underlying mechanism.

\begin{table*}[htbp]

         \renewcommand\arraystretch{1.25}

  \begin{tabular*}{0.98\hsize}{@{}@{\extracolsep{\fill}}cccccc@{}}
  \toprule[1pt]
  Type & $J_{i}~(K_{i}^{\pi})$  & $J_{f}~(K_{f}^{\pi})$ & Exp  & Refs.~\cite{nikolay_2017_PRL, nikolay_2019_PRL, nikolay_2022_PRC}   & PSM \\
  \midrule[0.6pt]
  \multirow{7}{*}{$E2$}  & $9/2~(5/2^{+})$   & $7/2~(5/2^{+})$   & $170 \pm 30$  & $213~(224)$       &  $191.5$ $(239.8)$   \\
        & $9/2~(5/2^{+})$   & $5/2~(5/2^{+})$  & $65 \pm 7$    & $82~(85)$          & $66.4$ $(83)$     \\
        & $9/2~(5/2^{+})$   & $5/2~(3/2^{+})$  & $6.2 \pm 0.8$ & $19.98~(17.37)$    & $4.5$ $(5.6)$    \\
        & $7/2~(5/2^{+})$   & $5/2~(5/2^{+})$  & $300 \pm160$  & $252~(267)$        & $233.0$ $(291.6)$    \\
        & $5/2~(3/2^{+})$   & $5/2~(5/2^{+})$  & $\cdots$      & $27.11$ - $39.49$~\cite{nikolay_2022_PRC}     & $8.4$ $(10.6)$   \\
        & $5/2~(3/2^{+})$   & $3/2~(3/2^{+})$  & $\cdots$      & $234.86$ - $239.18$~\cite{nikolay_2022_PRC}    & $222.2$ $(278.1)$   \\
        & $3/2~(3/2^{+})$   & $5/2~(5/2^{+})$ & $\cdots$      &  $27.04~(23.05)$   & $8.74$ $(10.95)$  \\
        \midrule[0.6pt]      
  \multirow{9}{*}{$M1$} & \multirow{2}{*}{$9/2~(5/2^{+})$}  & \multirow{2}{*}{$7/2~(5/2^{+})$}  & \multirow{2}{*}{$0.0076 \pm 0.0012$}  & $0.0178~(0.0157)$     & $0.0074$  \\ [-1.0ex]
        & & & & $0.0038$ - $0.0185$~\cite{nikolay_2019_PRL} &($0.0011$, $0.0015$) \\
        & \multirow{2}{*}{$9/2~(5/2^{+})$}  & \multirow{2}{*}{$7/2~(3/2^{+})$}  & \multirow{2}{*}{$0.0117 \pm 0.0014$}  & $0.0151~(0.0130)$       & $0.0085$    \\ [-1.0ex]
        & & & & $0.0144$ - $0.0151$~\cite{nikolay_2019_PRL} &($0.0143$, $0.0108$) \\
        & \multirow{2}{*}{$7/2~(5/2^{+})$}  & \multirow{2}{*}{$5/2~(5/2^{+})$}  & \multirow{2}{*}{$0.011 \pm 0.004$}  & $0.0093~(0.0085)$         & $0.005$    \\ [-1.0ex]
        & & & & $0.0011$ - $0.0096$~\cite{nikolay_2019_PRL} & ($0.0005$, $0.001$)\\
        & \multirow{2}{*}{$5/2~(3/2^{+})$}  &\multirow{2}{*}{$5/2~(5/2^{+})$}  & \multirow{2}{*}{$0.00326 \pm 0.00076$~\cite{nikolay_2022_PRC} }         & \multirow{2}{*}{$0.0012$ - $0.0050$~\cite{nikolay_2022_PRC}}          &$0.00339$   \\[-1.0ex]
        & & & & &($0.0015$, $0.0018$)\\
        & \multirow{2}{*}{$5/2~(3/2^{+})$}  & \multirow{2}{*}{$3/2~(3/2^{+})$}  & \multirow{2}{*}{$0.0318_{-0.0091}^{+0.0102}$~\cite{nikolay_2022_PRC}}    & \multirow{2}{*}{$0.0332$ - $0.0648$~\cite{nikolay_2022_PRC}}           & $0.0174$    \\ [-1.0ex]
        & & & & &($0.0288$, $0.0185$)\\
        & \multirow{3}{*}{$3/2~(3/2^{+})$} & \multirow{3}{*}{$5/2~(5/2^{+})$} & $0.0172^{+0.0031}_{-0.0023}$~\cite{kraemer_2023_nature} $0.0219^{+0.0006}_{-0.0006}$~\cite{tiedau_2024_PRL}   & \multirow{3}{*}{$0.0076~(0.0061)$}  & \multirow{3}{*}{ $0.0240$}    \\  
        & & & $0.0272^{+0.0074}_{-0.0082}$~\cite{yamaguchi_2024_nature}  $0.0295^{+0.0013}_{-0.0012}$~\cite{elwell2024laser}& &\\ 
        & & & $0.0213^{+0.0013}_{-0.0012}$~\cite{Hiraki_2024_NC}  $0.0214^{+0.0002}_{-0.0001}$~\cite{zhangck_2024_nature}& $0.0056$ - $0.0081$~\cite{nikolay_2019_PRL} &($0.0274$, $0.0226$)\\ 
  \bottomrule[1pt]
  \end{tabular*}
\caption{The reduced $E2$ and $M1$ transition probabilities (in W.u.) obtained from experiments \cite{NNDC,kraemer_2023_nature,tiedau_2024_PRL,yamaguchi_2024_nature,elwell2024laser,masuda_2019_nature,Hiraki_2024_NC,zhangck_2024_nature}, Refs.~\cite{nikolay_2017_PRL,nikolay_2019_PRL,nikolay_2022_PRC}, and our PSM calculation. The experimental data without indicating references is adopted from NNDC \cite{NNDC}. Values from Refs.~\cite{kraemer_2023_nature,tiedau_2024_PRL,yamaguchi_2024_nature,elwell2024laser,Hiraki_2024_NC,zhangck_2024_nature} are derived from the measured half-life/lifetime of $^{229m}$Th assuming $M1$ is the only contribution. In the result of Refs.~\cite{nikolay_2017_PRL,nikolay_2019_PRL,nikolay_2022_PRC}, data without indicating references represents the result from two sets of parameters in Ref.~\cite{nikolay_2017_PRL}. For E2 transition values from PSM, the values inside (outside) the parentheses correspond to the results with the standard (modified) effective charges. For M1 transitions in PSM, the values outside the parentheses correspond to the results with the fitted spin gyromagnetic factors, while the values inside the parentheses are the two sets of results with bare spin gyromagnetic factors (left) and spin gyromagnetic factors quenched by 0.85 (right) for both neutrons and protons.}
      \label{Tab:transition}
  \end{table*}


\begin{figure*}[t]
  \begin{center}
    \includegraphics[width=0.95\textwidth]{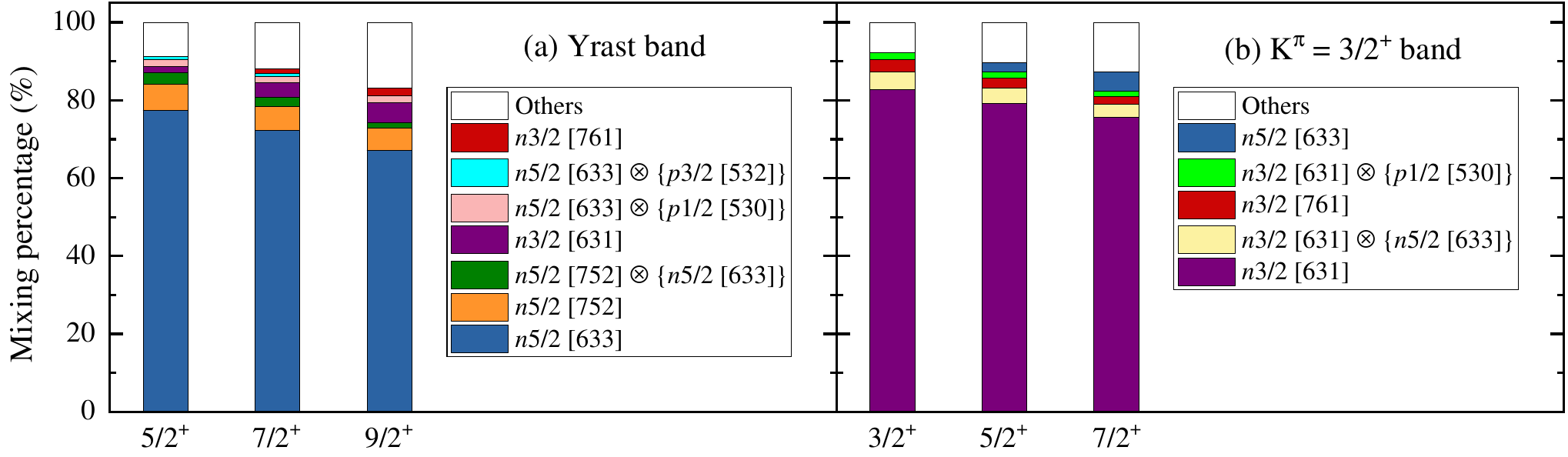}
    \caption{\label{fig:fig_2} The calculated configuration distributions of wave functions for states in the yrast and $K^\pi=3/2^+$ isomer bands by the PSM. The Nilsson single-particle label $\Lambda [N n_z m]$ is adopted, where $\Lambda$ is the total angular momentum projection along $z$ axis, $N$ denotes the major shell number, $n_{z}$ is the number of quanta in the $z$ direction, and $m$ is the orbital angular momentum projection along $z$ axis \cite{greiner_book, Ring_book_2004}. The Nilsson label in the brace represents coupled 2-qp state based on specific 1-qp state and its time reversal state, i.e., \{$\Lambda [Nn_{z}m]$\} stands for ($\Lambda [Nn_{z}m]) \otimes (-\Lambda [Nn_{z}m]$).}  
   \end{center}
\end{figure*}

The electromagnetic transition strengths play important roles for understanding the nature of $^{229m}$Th. We present in Table~\ref{Tab:transition} the reduced transition probabilities $B(E2)$ and $B(M1)$ for low energy transitions with the initial state $J_{i}~(K_{i}^{\pi})$ and final state $J_{f}~(K_{f}^{\pi})$. Two sets of effective charges: standard $0.5e$ ($1.5e$) and modified effective charge $0.6e$ ($1.6e$) \cite{Sheikh_PLB_2012} are used for neutrons (protons) to study $E2$ transitions. The numerical results are presented as pairs, where the values inside (outside) the parentheses correspond to the calculation with standard (modified) effective charge, respectively. For $M1$ transitions, the orbital gyromagnetic factors are taken from the truncated space \cite{JQChen_NPA_1998}. The values outside the parenthese are the results with the bare spin gyromagnetic factor of neutrons and the spin gyromagnetic factor of protons quenched by 0.85 which are fitted in the range between 0.7 and 1.0 by the known experimental transition probabilities of the low energy transitions excluding the direct decay of $^{229m}$Th. The two sets of values in the parentheses correspond respectively to the bare spin gyromagnetic factors and that quenched by 0.85, applied to both neutrons and protons. It is shown that $B(E2)$ values are large for in-band transitions but small for cross-band transitions, for states of the yrast and isomer bands. This behavior is in agreement with Ref.~\cite{nikolay_2017_PRL}. This is due to the fact that $B(E2)$ values are sensitive to the collective degrees of freedom in nuclear wave functions, indicating that the collective degrees of freedom are well described by the PSM wave functions.

We present in Fig. \ref{fig:fig_2} the configuration distributions of the PSM wave functions for the yrast and isomer bands, showcasing a visualization that enables the elucidation of the aforementioned phenomenon. The $5/2^+$ ground state exhibits a predominant occupation ($77.4\%$) of the 1-qp $n 5/2 [633]$ configuration with small contributions from other states, while the $3/2^+$ isomer shows strong configuration dominance (82.8\%) of the $n 3/2 [631]$ orbital. Their tiny configuration overlap, particularly through the shared $n 3/2 [631]$ component, results in small $B(E2)$ values for cross-band transitions due to attenuated transition matrix elements between these distinct nuclear configurations. The decay of $^{229m}$Th to the ground state is determined by $E2$ and $M1$ transitions. As shown in Table \ref{Tab:transition}, the corresponding $B(E2)$ from the PSM calculation is $8.74$ ($10.95$) W.u. which is smaller than the one in Ref.~\cite{nikolay_2017_PRL}, indicating that the $E2$ channel may probably be less important for the decay of $^{229m}$Th than one expected from Ref.~\cite{nikolay_2017_PRL}. Furthermore, as shown in Fig.~\ref{fig:fig_2}, along either the yrast or the isomer band, nuclear states with different angular momenta have similar collectivity with small changes of their wave functions. This explains the large in-band $B(E2)$ values presented in Table~\ref{Tab:transition}.



Due to the small $B(E2)$ value for the isomeric transition and its small excitation energy, the radiative decay of $^{229m}$Th is expected to be dominated by the $M1$ transition. Different from $E2$ transitions, for odd-mass nuclei, $M1$ transition strengths are more sensitive to the details of the nuclear wave functions, especially the configuration structure (single-particle or qp degrees of freedom), than to the collectivity. The three $B(M1)$ values from the NNDC database, as presented in Table \ref{Tab:transition}, for both in-band and cross-band transitions for states $9/2^{+}$ and $7/2^{+}$ of the yrast band are as small as $\lesssim0.015$ W.u.. This is well described by our PSM calculation, indicating that the single-particle degrees of freedom (configuration structure and mixing) are well described by the PSM wave functions shown in Fig. \ref{fig:fig_2}. 

Both $B(E2)$ and $B(M1)$ values for the transition from $^{229m}$Th to the ground state are so far unavailable in the NNDC database. However, one can derive the $B(M1)$ value from the recently measured half-life/lifetime of the isomer \cite{kraemer_2023_nature, tiedau_2024_PRL, yamaguchi_2024_nature, elwell2024laser, zhangck_2024_nature, Hiraki_2024_NC} by assuming the $M1$ transition is the only contribution for the radiative decay of $^{229m}$Th. These results differ from each other by about a factor of two, from $\sim 0.017$ to $\sim 0.03$ W.u., as shown in Table~\ref{Tab:transition}. This may indicate that further experimental and theoretical studies are desired. Refs.~\cite{nikolay_2017_PRL,nikolay_2019_PRL} predict smaller $B(M1)$ value than these experimental results. However, the corresponding $B(M1)$ from the PSM calculation is about $0.0240$ W.u., which is consistent with the recent measurements based on CaF$_2$ crystal \cite{ Hiraki_2024_NC, tiedau_2024_PRL, zhangck_2024_nature}. 

As shown in Eq. (\ref{eq.ROBTD}), $B(M1)$ can be calculated by summing over terms with different orbitals that consist of the single-particle matrix elements of the $M1$ operator and the ROBTD. The former is purely related to the $M1$ operator while the latter depends only on nuclear many-body wave functions. From our PSM calculations, the ROBTD is found to be very dispersed and turns out to be sizable not only for orbitals near the Fermi surface but also for orbitals far from the Fermi surface. This indicates that large model space is indispensable for reliable calculation of the decay of $^{229m}$Th. 

In addition, Ref.~\cite{masuda_2019_nature} reported the measured cross-band radiative width and Ref.~\cite{Sikorsky_2020_PRL} reported the radiative branching ratio for the second-excited state. As shown in Table~\ref{Tab:transition}, our result for the second-excited state agrees with these measured cross-band radiative width and radiative branching ratio. However, Ref.~\cite{masuda_2019_nature} reported also a large internal conversion coefficient for the cross-band transition and a total branching ratio between the in-band and cross-band transitions which is significantly different from the radiative one, indicating an implausible multipole mixing for the cross-band transition \cite{nikolay_2022_PRC}. Our calculation supports a small multipole mixing for the cross-band transition of the second-excited state as in Ref.~\cite{nikolay_2022_PRC}, suggesting that further investigations on the inconsistencies in the decay of the second-excited state \cite{masuda_2019_nature,Sikorsky_2020_PRL,nikolay_2022_PRC} should be necessary.

\section{Conclusions}

We have studied in the present work the nuclear structure of $^{229}$Th by the microscopic model of state-of-the-art projected shell model. In order to do this, we have developed here state-of-the-art PSM to provide a natural description of heavy odd-mass nuclei, with large model and configuration spaces, octupole degree of freedom and corresponding parity projection, as well as the involvement of most of the important interactions. Our calculation describes well low-energy levels of $^{229}$Th, and provides a reduced transition probability $B(M1)$ of $0.0240$ W.u. for the isomeric transition which agrees well with the radiative lifetime of $^{229m}$Th measured recently. Our calculation also supports a small multipole mixing for the cross-band transition of the second-excited state of $^{229}$Th. The physics behind these results has been revealed by analyzing the PSM nuclear wave functions.

The PSM calculation in the present work offers a deep insight into the unique $^{229}$Th from the microscopic nuclear structure point of view. The nuclear wave functions constructed here have been shown to be able to describe the nuclear properties of the $^{229}$Th nucleus, offering the chance of further studies for nuclear clocks and relevant topics by the nuclear wave functions. By the examination of the PSM predictions with the measured values for $^{229}$Th, which is a typical representative of heavy nuclei with an extremely low-energy isomeric state, the PSM has been shown to be a powerful method to study heavy nuclei. This might offer us the chance to study heavy nuclei in the unexplored regime.


\section*{Acknowledgements}

We thank Y. Sun and J. M. Yao for many valuable discussions. This work is supported by the National Natural Science Foundation of China (Grants No. 12475122 and No. 12275225), and by the Fundamental Research Funds for the Central Universities (Grant No. 010-63253121).

\appendix
\section*{Appendix A. Results with up to 5-qp configurations}

In Table~\ref{Tab:5qp-transition}, we present the results of the reduced electromagnetic transition strengths, $B(E2)$ and $B(M1)$, of $^{229}$Th calculated by the PSM with the configuration space up to 5-qp compared with the ones with configuration space up to 3-qp. For the $B(E2)$ values, standard effective charges are adopted. For the $B(M1)$ values, the bare spin gyromagnetic factor of neutrons and the spin gyromagnetic factor of protons quenched by 0.85 are adopted.

\setcounter{table}{0}
\renewcommand{\thetable}{A\arabic{table}}

\begin{table*}[h!]
 \caption{The reduced $E2$ and $M1$ transition probabilities (in W.u.) calculated by the PSM with different constrained configuration spaces where up to 3-qp and up to 5-qp configurations are considered respectively. } 
   \label{Tab:5qp-transition}
  \renewcommand\arraystretch{1.25}
    \begin{tabular*}{0.98\hsize}{@{}@{\extracolsep{\fill}}ccccccc@{}}
    \toprule[1pt]
    Type & $J_{i}~(K_{i}^{\pi})$  & $J_{f}~(K_{f}^{\pi})$    & PSM (3-qp)   & PSM (5-qp) \\
    \midrule[0.6pt]
    \multirow{7}{*}{$E2$}     & $9/2~(5/2^{+})$   & $7/2~(5/2^{+})$     &$191.5$     & $193.6$        \\
                              & $9/2~(5/2^{+})$   & $5/2~(5/2^{+})$     &$66.4$      & $66.3$         \\
                              & $9/2~(5/2^{+})$   & $5/2~(3/2^{+})$     &$4.5$       & $2.9$          \\
                              & $7/2~(5/2^{+})$   & $5/2~(5/2^{+})$     &$233.0$     & $233.8$        \\
                              & $5/2~(3/2^{+})$   & $5/2~(5/2^{+})$     &$8.4$       & $6.6$          \\
                              & $5/2~(3/2^{+})$   & $3/2~(3/2^{+})$     &$222.2$     & $223.3$        \\
                             & $3/2~(3/2^{+})$    & $5/2~(5/2^{+})$     &$8.74$      & $6.9$          \\
    \midrule[0.6pt]
        Type & $J_{i}~(K_{i}^{\pi})$  & $J_{f}~(K_{f}^{\pi})$ & PSM (3-qp)    &PSM (5-qp)\\
    \midrule[0.6pt]  
    \multirow{6}{*}{$M1$}    & $9/2~(5/2^{+})$  & $7/2~(5/2^{+})$  & $0.0074$        & $0.0071$\\ [-0.5ex]
                             & $9/2~(5/2^{+})$  & $7/2~(3/2^{+})$  & $0.0085$        & $0.0070$  \\ [-0.5ex]
                             & $7/2~(5/2^{+})$  & $5/2~(5/2^{+})$  & $0.005$         & $0.005$   \\ [-0.5ex]
                             & $5/2~(3/2^{+})$  & $5/2~(5/2^{+})$  & $0.00339$       & $0.00305$  \\
                             & $5/2~(3/2^{+})$  & $3/2~(3/2^{+})$  & $0.0174$        &$0.0188$   \\ 
                             & $3/2~(3/2^{+})$  & $5/2~(5/2^{+})$  & $0.0240$        & $0.0218$  \\  
    \bottomrule[1pt]
    \end{tabular*}
\end{table*}

\bibliographystyle{apsrev-no-url-issn.bst}
\bibliography{229Tharx.bib}{}

\end{document}